\documentstyle[prl,aps,epsf]{revtex}
\begin{document}
\title{Transient Hamiltonian chaos in the cavity electrodynamics }
\author{Kirill~N.~Alekseev${}^{(a)}$\cite{email1} 
and Gennady~P.~Berman${}^{(b)}$}
\address{
${}^{(a)}$Theory of Nonlinear Processes Laboratory,\\
Kirensky Institute of Physics, Krasnoyarsk 660036, Russia\\
${}^{(b)}$Center for Nonlinear Studies, Los Alamos National Laboratory,
Los Alamos, New Mexico 87545, USA\\}
\maketitle
\begin{abstract}
The chaotic dynamics at the interaction between an ensemble of two-level atoms with
an eigenmode of a high-$Q$ Fabry-Perot cavity and with an external
 amplitude-modulated
field is considered. It is shown that in the case of an exact atom-field
resonance and at the initial population of the atomic ground state,
a Hamiltonian chaos
in the system is transient. The difference in the chaotic dynamics for
a Fabry-Perot and for a ring cavity geometries, as well as the conditions for the experimental observation of the transient
Hamiltonian chaos are discussed.
\end{abstract}
\section{Introduction}
\par
 Tavis-Cummings (TC) model [1] which describes within the rotating-wave
approximation (RWA) [2] the interaction between an ensemble of
the two-level atoms with a high$-Q$ cavity mode is a fundamental for quantum and
nonlinear optics. Recently the basic dynamic effects predicted for
the model have become available for the experimental verification (see
[3,4] and references therein).
\par
Belobrov, Zaslavsky and Tartakovsky [5] showed that within the semiclassical
approximation a breakdown of the RWA leads to the dynamical chaos (see also [6]).
However, for the typical transitions in the optical and in the microwave domains,
the violation of the RWA may occur only at rather high atom's density, when a gas model
of the noninteracting two-level atoms is not more adequate.
\par
Nowadays other generalizations of the semiclassical
TC-model with the Hamiltonian chaos in the RWA appeared [7-13]. These models
describe the interaction of the three-level atoms with two cavity modes
[7,8], and two-level atoms with one mode
 and with an external field injected into a cavity with a constant amplitude
\footnote[1]{\noindent Note that this model may also describe the interaction of
impurity center in a crystal with the light and phonons [14]. In such
case a cavity is absent.}
  [9-11], or with the amplitude-modulated
 field [12]. For the last two cases, the interaction with
an external field was taken into account within a standard spatially -
homogeneous field approximation (``mean field model'') [15].
\par
The common feature of all mentioned above models is the consideration
of the sample shorter then the wavelength of the ring cavity with the only
forward propagating wave. For the Fabry-Perot cavity, an account of
spatial variation of the cavity field amplitude due to the
interference between the forward and backforward propagating waves may
influence sufficiently on the nonlinear characteristics of the atoms-field
interaction. This fact has been well appreciated in the theory of
optical bistability (see [16, 17] and references therein).
\par
In this paper we are concerned within a semiclassical approach
the influence of the standing wave effects on the transition to the Hamiltonian
chaos at the interaction between the two-level atoms with the field. We use
the model [12], but instead of a ring cavity a Fabry-Perot cavity
geometry is considered.
\par
It is shown, that in the simplest case of an exact atom-field
resonance and at the initial population of the atomic ground state, the
coupled Maxwell-Bloch system is reduced to the Hamiltonian system
with 1.5 degrees of freedom --- a periodically forced ``Bessel
pendulum''. Conditions for the transition to the dynamical chaos are found
numerically. For the majority of the values of the external
field amplitude and frequency,  chaos is transient: the chaotic
oscillations of the population
difference and of the polarization become regular after some time.
In contrast to other models of physical systems possessing
the transient Hamiltonian chaos (see, e.\ g.\  [18 - 23]), in our case the
effective potential of the Hamiltonian system with 1.5 degrees of
freedom is of a multi-well type. This leads to rather complicated nonlinear
dynamics of the system. In particular, a time interval of chaotic
behavior has complicated dependence on the system's parameters. For
trajectories with rather long chaotic part, a sensitivity of
a character of settling of a regular oscillations of a cavity field with respect to  small
variations of the initial conditions or the parameters of the system is found.
\par
The paper is organized as follows. In Sec.\ 2 we introduce the
model describing the dynamics of the system  ``two-level atoms +
field". The anzatz reducing the Maxwell-Bloch system to one
equation of periodically driven ``Bessel pendulum'' in the case of
the exact resonance and for some initial conditions is also considered in
this Section. In Sec.\ 3 after a brief review on
the trasient Hamiltonian chaos, we describe
the nonlinear dynamics of our system. In conclusion, in Sec.\ 4 we compare
the particularities in the dynamics of atoms-field interaction for
a geometry of the Fabry-Perot and ring cavities, and discuss the possibilities of
the experimental observation of the transient Hamiltonian chaos, and also point
out some open problems.
\section{Model}
\par
Our system consists of $N$ identical two-level atoms with
a transition frequency $\omega _{0}$ and a dipole matrix element $d$, interacting with
the radiation field mode with the frequency $\omega  \approx  \omega _{0}$ . The sample with atoms of
the length $L$ and the volume $V$ completely fills a high$-Q$ Fabry-Perot cavity
$(-L \leq  z \leq  0)$. An external amplitude-modulated field $E_{ext}$ is injected
into the cavity at the plane $z = 0.$ The field has the form
\begin{equation}
E_{ext} = E_{0} F (t) \cos  \omega  t ,
\end{equation}
where $F (t)$ is a periodic function of time $( \mid  F (t) \mid  = 1 )$ slowly
varying in comparison with the carrier frequency $\omega $. In this paper we
shall consider the simplest type of modulation $F (t) = \sin  \Omega  t$, but it can be
expected that the main results are also applicable for the case,
when the modulation $F (t)$ consists of many harmonics. Following
the classical paper of Spencer and Lamb [24], we shall consider the
influence of an external field as a small local perturbation preserving
a cavity mode structure of the form
\begin{equation}
E_{SC} (z, t) = E (t) \sin  (k z),
\end{equation}
where $k = n \pi  L^{-1}$ is the wave number of the chosen $n$-th cavity mode. The
dynamics of atoms is described by the Bloch equations
\begin{eqnarray}
\dot{S}_{1} (z, t) = - \omega _{0} S_{2} (z ,t), \nonumber \\
\dot{S}_{2} (z, t) = \omega _{0} S_{1} (z, t) + {2d \over \hbar} E_{SC} (z, t) S_{3} (z, t), \\
\dot{S}_{3} (z, t) = - {2d \over \hbar} E_{SC} (z, t) S_{2} (z, t) , \nonumber
\end{eqnarray}
\begin{equation}
S^{2}_{1} + S^{2}_{2} + S^{2}_{3} = 1 ,
\end{equation}
where the pseudospin variables $S_{i}$ $(i=1,2,3)$ are connected with
the probability amplitudes for the population of upper $a_{j}$ and lower $b_{j}$
levels of the $j$-th atom [9, 12, 25]
\begin{eqnarray}
S_{1}( z, t) = {1\over N_{s}} \Sigma^{N_{s}}_{j \in \Delta V} ( a^{*}_{j} b_{j} + a_{j} b^{*}_{j} ), \nonumber \\
S_{2}( z,t ) = - {i\over N_{s}} \Sigma^{N_{s}} _{j \in \Delta V} ( a^{*}_{j} b_{j} - a_{j} b^{*}_{j} ), \\
S_{3}(z,t) = {1\over N}_{s} \Sigma^{N_{s}} _{j \in \Delta V} ( \mid  a^{2}_{j} \mid  - \mid  b^{2}_{j} \mid  ). \nonumber
\end{eqnarray}
In (5) $\Delta V = (\Delta z) \pi  r^{2}$ is a small averaging volume, $z$ is the
coordinate of slice center with the width $\Delta z \ll  \lambda$  ($\lambda $ is wavelength), $r$
is a characteristic radius of the sample with a gas of atoms, $N_{s}$ is
the number of atoms in the volume $\Delta V$ ($N_{s} \gg  1$).
\par
The behavior of the field inside the cavity is governed by the Maxwell
equation including an influence of the external field [24]
\begin{eqnarray}
\ddot{E} ( t ) + \omega ^{2} E ( t ) = -4 \pi  \ddot{\tilde{P}} ( t ) - {2 \omega  c\over L} E_{\hbox{ext}}, \\
\tilde{P} ( t ) = {2\over L} \int ^{o}_{-L} dz P ( z, t ) \sin  ( k z ) , \nonumber
\end{eqnarray}
where $P(z,t)$ is a polarization of the two-level medium determined by
\begin{equation}
P (z, t) = {N\over V} d S_{1} (z, t).
\end{equation}
The coupled Maxwell-Bloch system (3), (6), (7) completely describes
the dynamics on the time interval shorter then all characteristic
relaxation times of atoms and field. These equations may be
simplified by separation of fast and slow dynamics. For this purpose
make a transformation into the rotating system of coordinates
\begin{eqnarray}
S_{1} (z, t) = u (z, t) \cos  (\omega  t) - v (z, t) \sin  (\omega  t), \nonumber \\
S_{2} (z, t) = u (z, t) \sin  (\omega  t) + v (z, t) \cos  (\omega  t), \\
S_{3} (z, t) = w (z, t), \nonumber
\end{eqnarray}
and introduce the envelopes of the self-consistent field in the following way
\begin{equation}
E ( t ) = E_{1} ( t ) \cos  ( \omega  t ) + E_{2} ( t ) \sin  ( \omega  t ).
\end{equation}
Assume that the conditions of a slow variance of the envelopes in a comparison with
the carrier frequency $\omega $
\begin{eqnarray}
\mid  \dot{u} \mid  \ll  \omega  \mid  u \mid  , \quad  \mid \dot{v} \mid  \ll  \omega  \mid  v \mid  , \quad \mid  \dot{w} \mid  \ll  \omega  \mid  w \mid  , \nonumber \\
\mid  \dot{E}_{1,2} \mid  \ll  \omega  \mid  E_{1,2} \mid  ,
\end{eqnarray}
are valid. Then, substituting (1), (8) into (3), (6), (7) and using the RWA
and a slowly varying envelope approximation (SVEA) [2] one can
derive the following equations
\begin{eqnarray}
\dot{u}= \Delta  v + w {\cal E}_{2} \sin  ( k z ), \nonumber \\
\dot{v}= - \Delta  u + w {\cal E}_{1} \sin  ( k z ), \nonumber \\
\dot{w}= - (u {\cal E}_{2} + v {\cal E}_{1}) \sin  ( k z ) \\
\dot{{\cal E}}_{1} = \omega ^{2}_{c} {2\over L} \int ^{0}_{-L} dz v ( z, t ) \sin  ( k z ) + G F (t), \nonumber \\
\dot{{\cal E}}_{2} = \omega ^{2}_{c} {2\over L} \int ^{0}_{-L} dz u ( z, t ) \sin  ( k z ). \nonumber
\end{eqnarray}
In (11) $\Delta  = \omega  - \omega _{0}$ is a detuning from the optical resonance, $G = (c\epsilon _{0}) /
L$, ${\cal E}_{j} = d E_{j}/ \hbar$ $ (j=0, 1, 2)$, $\omega _{c} = (2 \pi  N d^{2} \omega _{0}/ \hbar V)^{1/2}$ is a
so-called cooperative frequency [26] characterizing the time of
energy transfer between atoms and field in the absence of the external
field $(E_{0} = 0)$.\footnote[2]{\noindent In modern literature also notions such as ``collective Rabi
frequency'' and ``vacuum- field Rabi frequency'' are used for the
determination of this frequency (see the discussion of
the corresponding terminological questions in [4]).}
The Maxwell-Bloch equations possess the pseudospin
length conservation law
\begin{equation}
u ( z, t )^{2} + v ( z, t )^{2} + w ( z, t )^{2} = 1.
\end{equation}
Substituting (8) into (7), we obtain the representation for
the polarization via the slow variables $u$ and $v$
\begin{equation}
P (z, t) = {Nd\over V} \left[ u (z, t) \cos  (\omega  t) - v (z, t) \sin  (\omega  t) \right].
\end{equation}
The condition of applicability of the RWA and the SVEA may be written in
the form
\begin{equation}
\max \left( G^{1/2}, \omega _{c}, \Omega \right) \ll  \omega  \sim \omega _{0}.
\end{equation}
As a rule, for typical optical and microwave systems this condition
is valid.
\par
Now the system (11) contains only slow variables. To solve it, it
is necessary to define the initial spatial distributions of
polarization components $u (z, t = 0) \equiv  u_{0} (z)$ and $v_{0} (z)$ as well as
population difference $w_{0} (z)$ and field envelopes ${\cal E}_{1,2} (0)$.
Deriving a general solution of (11) looks as rather
difficult problem, so we consider only particular solution for
an exact resonance and at the specific initial condition.
\par
It may be shown that for an exact resonance $( \Delta  = 0 )$, if
$u (z, t = 0) = {\cal E}_{2} (t = 0 ) = 0$, then $u (z, t) = {\cal E}_{2} (t ) = 0$ for any
$t$. \footnote[3]{Or, if
$v (z, t = 0) = {\cal E}_{1} (t = 0 ) = 0$, then $v (z, t) = {\cal E}_{1} (t ) = 0$
for any $t$.}
In this case it follows from (12) that the polarization $v (z, t)$ and
the population difference $w (z, t)$ may be parameterized by one angle
variable $\phi  (z, t)$
\begin{eqnarray}
v (z, t) = - \sin  \phi  (z, t), \nonumber \\
w (z, t) = - \cos  \phi  (z, t).
\end{eqnarray}
Introduce the following anzatz
\begin{eqnarray}
\phi  (z, t) = \vartheta  ( t ) \sin  (k z) , \\
\vartheta  ( t ) = \int ^{t}_{0} dt^{\prime}  {\cal E}_{1} ( t^{\prime}).\nonumber
\end{eqnarray}
From (15) one can derive
\begin{eqnarray}
v (z, t) = - 2 \Sigma ^{\infty }_{n=0} J_{2n+1} (\vartheta  (t) ) \sin  [ (2n+1) kz], \nonumber \\
w (z, t) = - J_{0} (\vartheta  (t) ) - 2 \Sigma ^{\infty }_{n=0} J_{2n} ( \vartheta  (t) ) \cos  (2n kz ) ,
\end{eqnarray}
where $J_{n} (x)$ is the Bessel function. In these notastions the equations (11) may be transformed to one equation
of the periodically driven ``Bessel pendulum''
\begin{eqnarray}
\ddot{\vartheta} + 2 \omega ^{2}_{c} J_{1} ( \vartheta  ) = G \sin  \Omega  t , \\
\dot{\vartheta} = {\cal E}_{1} \equiv  {\cal E} ( t ). \nonumber
\end{eqnarray}
Taking into account that for small $x$: $ J_{1} (x) \approx  x / 2$, one can see
that the cooperative frequency $\omega _{c}$ is just the frequency of small
oscillations of the pendulum at $G = 0$.
\par
As far as we know, the anzatz (16) , (17) was first introduced in
[27] when studying the metastable states in the coupled ``atoms +
field" system. The equation of the ``Bessel pendulum'' without the external driving $(
G = 0)$ has been also obtained in [27].
\par
Now we discuss  the problem of the limits of applicability of the anzatz
(16), (17). Expansion (17) should be correct for any $v (z, t)$ and $w (z, t)$
including the initial distribution $v_{0} (z)$ and $w_{0} (z)$. It means that
should exist such $\vartheta  (0) \equiv  \vartheta _{0}$ , for which the expansions
\begin{eqnarray}
v_{0} ( z, t) = - 2 \Sigma ^{\infty }_{n=0} J_{2n+1} ( \vartheta _{0} ) \sin [( 2n+1 ) kz], \nonumber \\
w_{0} (z, t) = -J_{0} ( \vartheta _{0} )- 2 \Sigma ^{\infty }_{n=0} J_{2n} ( \vartheta _{0} ) \cos (2n kz ) ,
\end{eqnarray}
are identities.\footnote[4]{\noindent Prof.\ Wei-Mou Zheng called our attention to this fact.}
But the fulfillment of (19) is possible not for
all $v_{0} (z)$ and $w_{0} (z)$. Among the initial distributions $v_{0} (z)$ and $w_{0}
(z)$ for which (19) is correct, there are at least two
physically significant cases: (i) the case of the periodic $\delta $-like
distribution and the population difference corresponding to $\vartheta _{0} \gg  1$, and
(ii) the case of spatially homogeneous and weak initial excitation
of two-level medium $w_{0} (z) \approx  -1$ and $v_{0} (z) \approx  0.$ Such initial
distribution corresponds to $\vartheta _{0} \approx  0$. Note that anzatz (16), (17)
can be used for any initial values of the field ${\cal E} (0)$.
\par
In what follows, we shall consider physically most interesting
case of atoms initially populated in the ground state ($ w_{0} (z) = -1$,
$v_{0} (z) = 0$), and when initially the field in the cavity is absent ($ {\cal E} (0) = 0
$). It means that for the pendulum (18), the initial conditions are
fixed: $\vartheta  (0) = \dot{\vartheta} (0) = 0$.
\section{Hamiltonian Dynamics}
\par
Equation (18) may be written in the Hamiltonan form
\begin{eqnarray}
{d\vartheta \over d\tau} = {\partial H\over \partial p} , \quad {dp\over d\tau} = - {\partial H\over \partial \vartheta }, \nonumber \\
H = p^2 /2 + V ( \vartheta ) - \overline{G} \vartheta  \sin (\overline{\Omega} \tau) , \\
V (\vartheta ) = - 2 J _{0} (\vartheta ) , \nonumber
\end{eqnarray}
where $p \equiv  {\cal E} / \omega _{c}$, $\tau = \omega _{c} t$, $\overline{G} = G  / \omega ^{2}_{c}$, $\overline{\Omega} = \Omega  / \omega _{c}$ . The form of
potential $V (\vartheta )$ is shown in Fig.~1.
\par
Transfer into extended phase space [28] and introduce a pair
of new canonically conjugated variables $(\psi , I)$  according to the
formulas
\begin{eqnarray}
{d\psi \over d\tau} = {\partial \tilde{H}\over \partial I} , \quad  {\hbox{dI}\over d\tau} = - {\partial \tilde{H}\over \partial \psi } ,  \\
\psi  = \overline{\Omega} \tau  \quad \tilde{H} = H + \overline{\Omega} I. \nonumber
\end{eqnarray}
The new Hamiltonian $\tilde{H}$ is the integral of motion in the extended phase space $(\vartheta ,
p, \psi , I)$. We used formulas (21) to control the errors in the numerical
calculations.
\par
In spite of rather simple form of the equation (18), the dynamics
of this Hamiltonian system with 1.5 degrees of freedom may be very
complicated. Such a behavior is due to the form of the potential $V (\vartheta )$
which is not periodic, and is a decreasing function of $\vartheta $ at $\mid  \vartheta  \mid  \to  \infty $ .
Under the influence of a periodic forcing both regular and chaotic
dynamics are possible. As $V (\vartheta ) \to  0$ at $\vartheta  \to  \pm  \infty $ , then the asymptotic
behavior of the dynamics is regular. Therefore, among with trapped periodic
and chaotic trajectories, transient chaos is possible.
\par
In recent years the investigations on the transient Hamiltonian chaos
attract considerable attention (see, e.\ g.\ , reviews [18,19] and
references therein). The majority of papers deal with the study of
particle scattering on $2D$ or $3D$ potentials [18-20]. It was quite
recently realized that stochastic ionization of atoms and molecules
or, generally speaking, any escape of chaotic trajectory from
the potential well are also examples of the transient Hamiltonian chaos.
The transient chaos under the conditions of stochastic ionization is not
only much more poor studied, but also is organized more complicated
then chaos in the potential scattering [22]. This is due to the fact
that the phase space structure of the typical Hamiltonian systems is
not uniform: Along with the stochastic seas the great number of
islands with the stable dynamics are presented, which are the source of
the trajectory trapping for a long time.
\par
In paper [21] the stochastic ionization  of kicked Morse
oscillator is considered. Initial oscillator's state corresponds
to the potential minimum. It was shown, that the time of chaotic escape
from the well $t_{esc}$ may be sensitive to a small change of the
parameters of the system. The boundaries of the regions with the different $t_{esc}$ are
fractals in the space of parameters: amplitude --- period of external
driving.
\par
In paper [23] the escape of chaotic trajectory from $2D$
potential has been studied. It was shown that for strong chaos the
directions and time of escape from the well are significantly sensitive to
a small change of the initial conditions. The region in the space of
the initial conditions corresponding to the trajectory escape during a fixed
number of iterations appears to be a fractal.
\par
All mentioned above papers were devoted to the study of chaotic escape
from $2D$ potential (2 degrees of freedom), or from $1D$ potential
which consists of one potential well (1.5 degrees of freedom). The particularity
of our system with 1.5 degrees of freedom consists in multi-well
structure of the effective potential (see Fig.~1). It should be noticed also, that
in our case initial conditions are fixed just in the bottom of the first well.
\par
There may exist several types of the trajectories (both trapped and
unbounded): stable periodic, unstable periodic, trapped chaotic and
unbounded chaotic demonstrating transient chaos. Due to nonlocal structure of
our effective potential, the trajectory can be trapped not only within the
first (central) well, but inside other wells too. Of course, time interval
of trajectory trapping inside different wells is quite different. Trapped
chaotic trajectories exist only for some singular values of parameters and
the most typical behavior is transient chaos.
During transient chaotic motion trajectory after escaping from the first
well may randomly visit several wells and be trapped within these wells
for some time. So, because of nonlocal character of the effective
potential, transient chaotic motion may be organized rather complicated.
\par
We illustrate now the main types of the nonlinear dynamics. First, if
the conditions of chaos are not satisfied, then a trajectory is located in
the first well $(-3.83 \leq  \vartheta  \leq  3.83)$ forever. A corresponding form of the field
(momentum of effective system (20)) for regular motion is shown in
Fig.~2. Under the conditions of chaos the behavior of the system is more
various:
\par
\noindent (a) An example of unstable quasiperiodic trajectory trapped inside
first (central) well during all time of computation is shown in Figs.~3,~4.
Such trajectories are observed mainly for
the parameter values belong the vicinity of the chaos boundary.
\par
\noindent (b)
Examples of transient chaos are shown in Figs.~5 and~6. The trajectory quickly
escapes from first well, visits randomly and ``rotates'' during some time inside
few other wells and, by the end, a regular asymptotic motion sets in. (Note
that minimums of the wells correspond to the values $ \vartheta \approx 0, \pm
7, \pm 13.3, \pm 19, \pm 25$ etc.).
The signs of the coordinate $\vartheta $
and the momentum $p$ setting in at $t \to  \infty $ are random. Very small variation of
the initial conditions leads to absolutely different asymptotic state
(compare Figs.~5 and 6). This results is the interesting physical
effect: the small fluctuations of the initial cavity field produce the
change of the asymptotic field state sign (Fig.~7).
\par
\noindent (c) An example of chaotic trajectory trapped during all time
of computation is shown in Fig.~8.
The
trajectory quickly escapes from the first well $( \tau_{esc} \approx  10.8 )$, then
visits few neighbor wells and again returns to the first well.
Corresponding chaotic field oscillations are shown in Fig.~9.
\par
\noindent (d) By the other hand, a transition to the regular behavior may be very
quick and just after escaping from first well (Figs.~10, 11).
\par
The regions of regular and chaotic behavior in the space of
parameters: amplitude of driving --- frequency of modulation are shown
in Fig.~12. The chaos boundary is marked by stars, the regularity
windows lying near this boundary are marked by boxes. The minimal
value of the external field amplitude $( \overline G \approx  0.14 )$ leading to chaos was
observed at the frequency $\overline \Omega \approx  0.85$ . The trajectories with chaotic parts
of different length exist in the regions situated above the chaos boundary.
These trajectories have positive maximal Lyapunov exponent during
the time interval of chaotic motion. Our preliminary numerical
simulations show rather complicated dependence of the length of the chaotic phase on
the parameters. But definitely the trajectories
with long chaotic path $( \tau_{esc} \sim 10^{2} )$, as a rule, belong to the layer
with the width $\sim 0.1$ near the boundary with the chaotic component.
\par
We also studied the sensitivity of the sign of the self-consistent
field asymptotic state with respect to the small variation of the external field
amplitude. The results are presented in Fig.~13. Dimensionless
perturbation parameter $\overline G$  was changing in the region [0.89 , 0.9] with
a step $10^{-4}$. A regular asymptotic state with a positive field $( {\cal E} > 0 )$
is displayed by plus one ( +1 ), and the opposite case of the negative field $(
{\cal E} < 0 )$ is marked by minus one ( -1 ). It is seen from Fig. 13,
that in some regions of the perturbation parameter values $\overline G$ , the sign
of the asymptotic field state is sensitive to small variation of $\overline G$.
Such kind of behavior is characteristic for trajectories with rather long
chaotic path $( \tau_{esc} \approx  40 \div  100)$.
\par
In concluding part of this section we compare spatial structure
of polarization and population difference at regular and chaotic
dynamics (see formula (17)). An essential contribution into the
expansion (17) make only spatial harmonics with the index
of the Bessel functions less then its argument. For regular
dynamics, the argument of Bessel function $\vartheta $ is smooth regular
function of time. In this case the spatial spectrum of $v$ and $w$ may
contain many harmonics, but the energy transfer between different
spatial modes is regular. By contrast, for chaotic temporal
dynamics $\vartheta $ is random function of time. As a consequence, spatial
chaos in the distribution of the population difference and polarization
appears.
\section{Discussion}
\par
It is interesting to compare a nonlinear dynamics of the system
of two-level atoms + self-consistent field + external amplitude-modulated
field for different geometry of the cavity. Generalized coupled
Maxwell-Bloch system may be reduced to the periodically driven
``Bessel pendulum'' (18) in the case of the Fabry - Perot cavity.
For the majority of paremeter values, Hamiltonian chaos is transient
in this system. By contrast, for
the ring cavity, the dynamics of the system within the same approximation
is governed by the equation of the periodically driven physical pendulum [12]
\[   \ddot{\vartheta} + \omega ^{2}_{c} \sin  \vartheta  = G F ( t ) ,  \]
where $\vartheta  (t)$ is the Bloch angle. Due to periodicity of the corresponding
potential $V (\vartheta ) = - \omega ^{2}_{c} \cos  \vartheta $ and for periodic $F (t)$, the dynamical chaos is
stationary. Appearance of the transient chaos is possible only for
a special choice of $F (t)$, i.\ e.\ when $F (t)$ contains zero harmonic
[12]. Then, the effective potential loses its translation invariance.
\par
At present time the oscillatory energy transfer between atoms and
field has been observed in both optical and microwave domains [4].
Present here some estimates. For dipole transition with $d \sim 1$Debye and
atom's density $10^{14} cm^{-3}$ , the value of the cooperative frequency $\omega _{c} \sim
10^{10} s^{-1}$ may be large then the dissipative rates associated with
the atom relaxation and cavity losses. We believe that the rapid progress in
the experimental cavity quantum electrodynamics makes observation of
the transient Hamiltonian chaos accessible.
\par
In summary, we considered the influence of the spatial structure of
a standing wave on the chaotic dynamics in the interaction of an ensemble
of two-level atoms with the field. It is shown, that for the
majority of paremeter values, chaos is transient.
For trajectories with long chaotic path, the asymptotic behavior of
the field is sensitive to small variation of the initial conditions and
parameters. Temporal chaos has manifestation in the spatial behavior of
the polarization and of the population difference.
\par
In conclusion we point out some relevant problems for future
research. First, a complicated nonlinear dynamics of periodically
forced ``Bessel pendulum'' deserves more detail and deep study in
the frameworks of general problem of transient chaos. Second, the
problem of influence of quantum effects on chaos in potential
scattering attracts great attention of researches (see review [29]).
But the influence of quantum effects on the other types of the transient
Hamiltonian chaos is yet not considered. So, it could be interesting to
study the influence of the cavity field quantalization on the transient
chaos in the model discussed.
\section*{Acknowledgements}
\par
We thank P.W. Milonni for discussion and V.\ S.\ Egorov for sending us 
a reprint of the paper
[27]. One of us, A.\ K.\ N.\ , also acknowledges Hao
Bai-Lin and Wei-Mou Zheng for hospitality, support and useful
discussions during his visit of ITP (Beijing), where a big part of
this work was done. 

\section*{Figure Captions}

{\em Figure 1}. The form of the potential for the Bessel pendulum $V (x) = - 2 J_{0}
(x)$.\\

{\em Figure 2}. Dependence of the self-consistent field $p \equiv  {\cal E}  / \omega _{c}$ on time
for regular motion inside the first potential well: $ \overline{G} = 0.8$;
$\overline{\Omega}  = 1.3$.\\

{\em Figure 3}. Weakly unstable chaotic orbit of length $\tau = 200$ located in
the first well: $\overline{G}  = 0.271$; $\overline{\Omega}  = 0.8$.\\

{\em Figure 4}. Dependence of the field on time. The values of parameters are the same as
in Fig.~3.\\

{\em Figure 5}. Trajectory of the length $\tau = 200$ under the conditions of the transient
chaos ($\overline{G}  = 0.7$; $\overline{\Omega}  = 0.6$).\\

{\em Figure 6}. Same as Fig.~5, but with slightly changed initial conditions for the field: $p (0) = 10^{-4}$.\\

{\em Figure 7}. Changing of the asymptotic field state at small variation of
the initial conditions. Process marked by boxes corresponds
to the initial condition $p (0) = 0,$ not marked process ---
$p (0) = 10^{-4}$. Values of parameters are the same as in Figs.~5 and
6.\\

{\em Figure 8}. Chaotic trajectory of length $\tau = 200$ at $\overline{G}  = 0.6$; $\overline{\Omega}  = 0.9$.\\

{\em Figure 9}. Chaotic field oscillations for the parameter values are the same as in
Fig.~8.\\

{\em Figure 10}. Quick escape from the first potential well $( \tau_{esc} = 14.8 )$
and setting of the regular motion at $\overline{G} = 0.4$; $\overline{\Omega} = 0.9$.\\

{\em Figure 11}. Dependence of the field on time. Values of parameters are the same as
in Fig.~10.\\

{\em Figure 12}. Regions of regular and chaotic motion in parameter space:
amplitude --- frequency.\\

{\em Figure 13}. The influence of the variation of the driving amplitude on the sign of
the asymptotic field state (for explanation, see main text),
$\overline{\Omega} = 0.4$.\\

\epsfxsize=10cm
\hspace{2cm}
\vspace{3cm}
\epsfbox{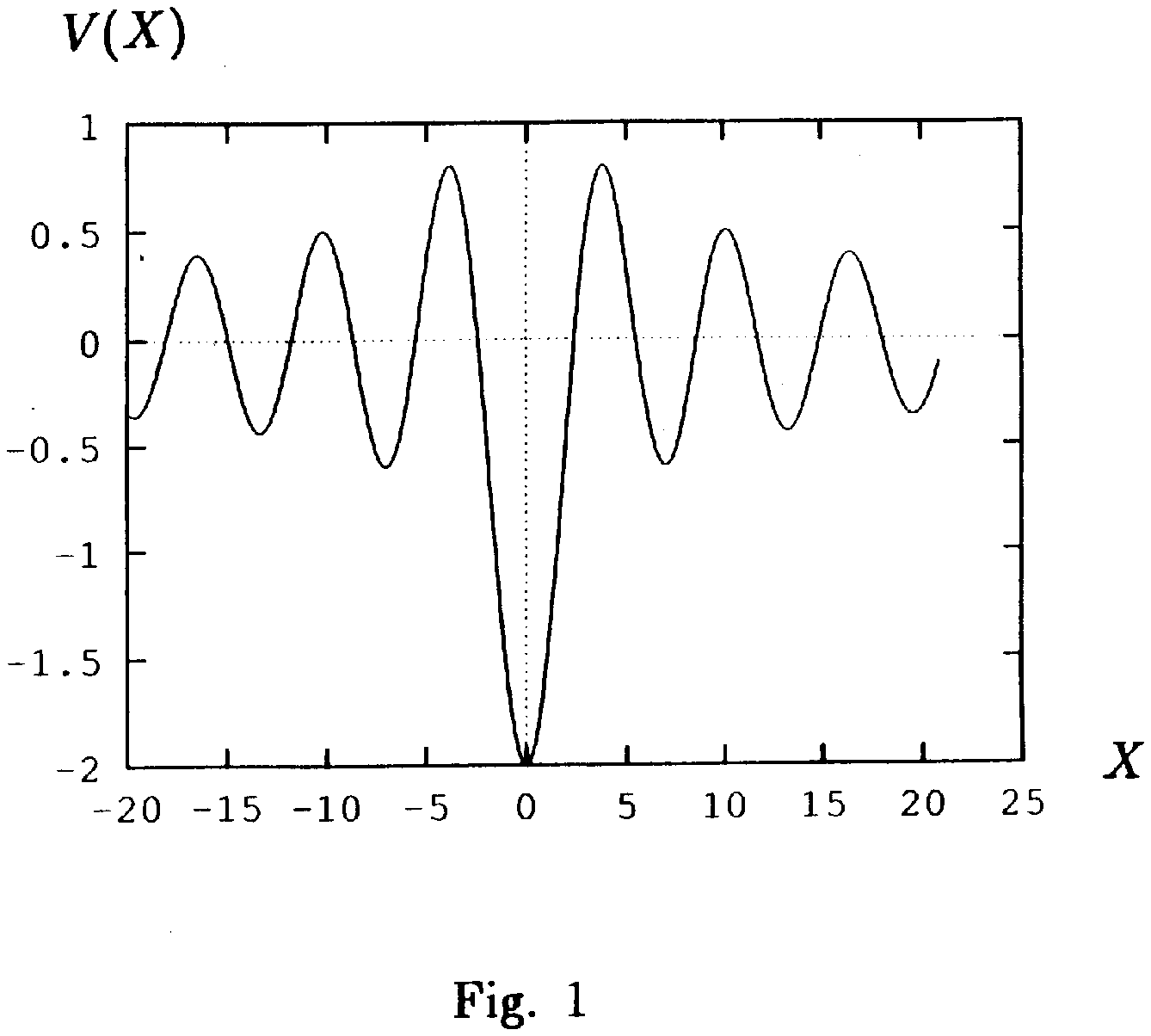}

\epsfxsize=10cm
\hspace{2cm}
\epsfbox{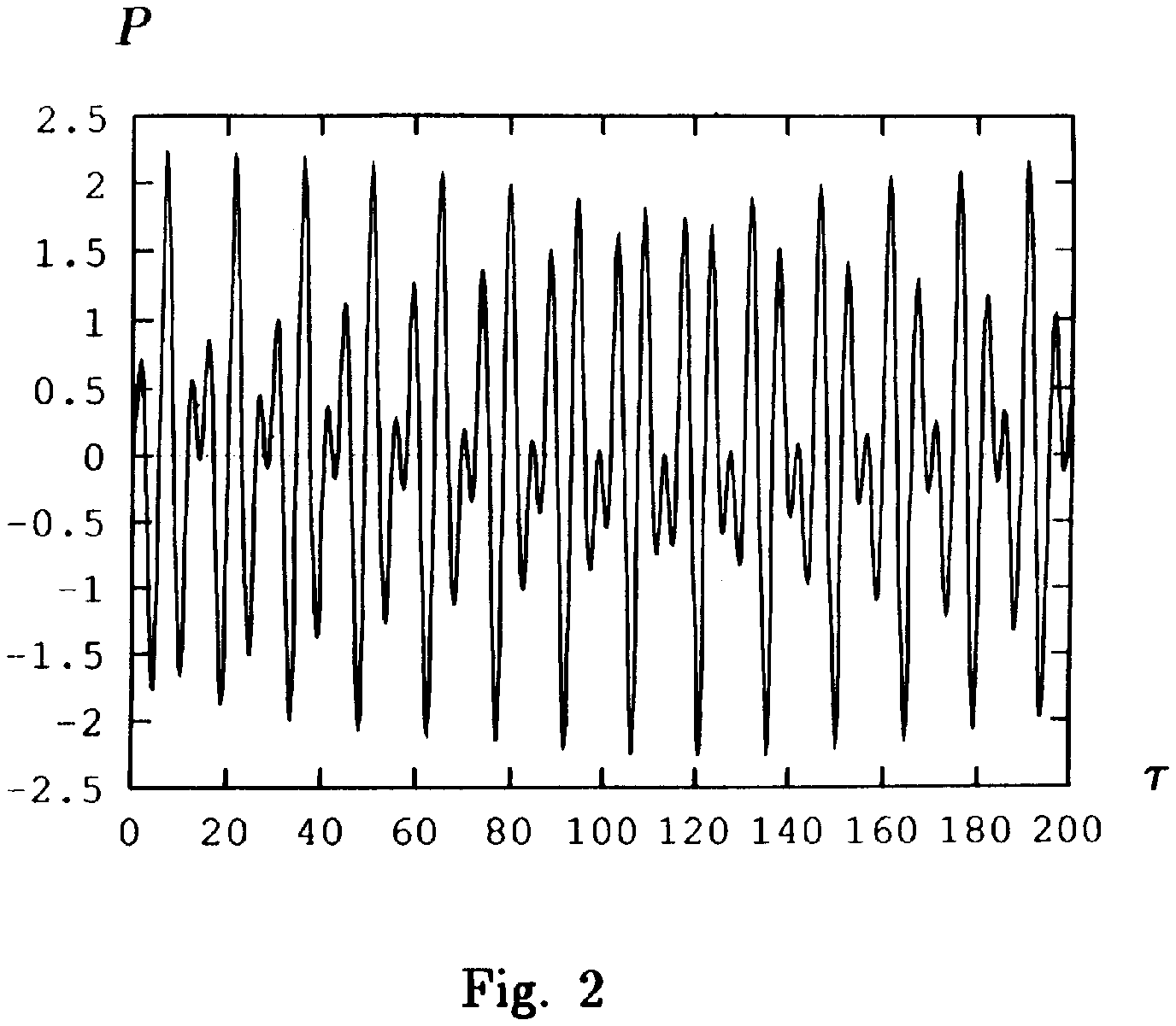}

\epsfxsize=10cm
\hspace{2cm}
\vspace{3cm}
\epsfbox{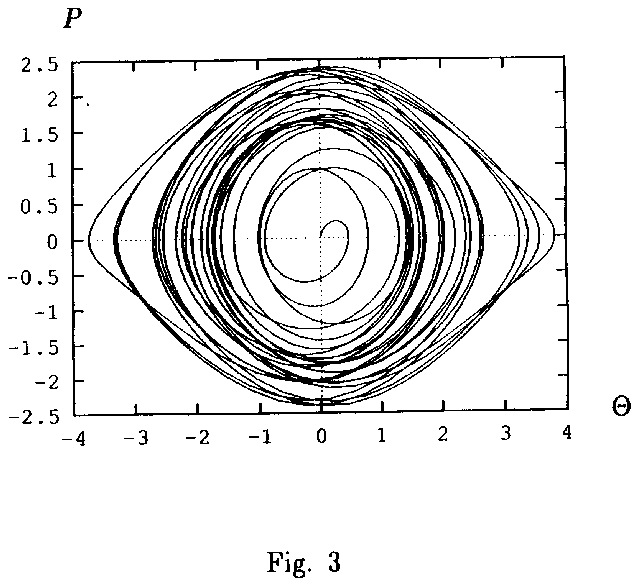}

\epsfxsize=10cm
\hspace{2cm}
\epsfbox{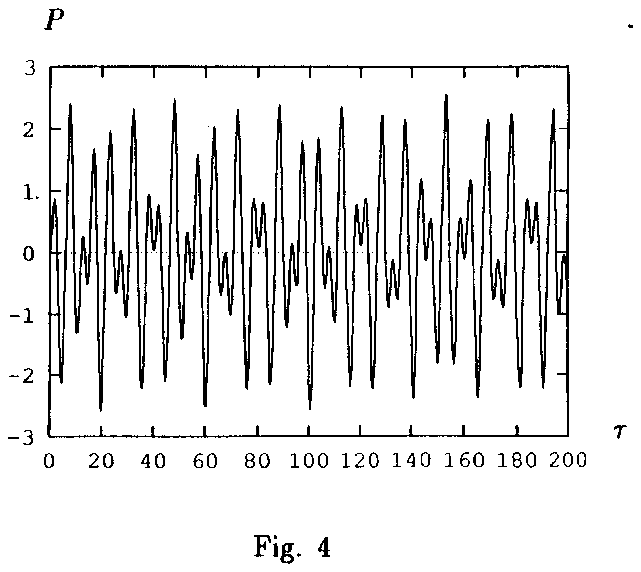}

\epsfxsize=10cm
\hspace{2cm}
\vspace{3cm}
\epsfbox{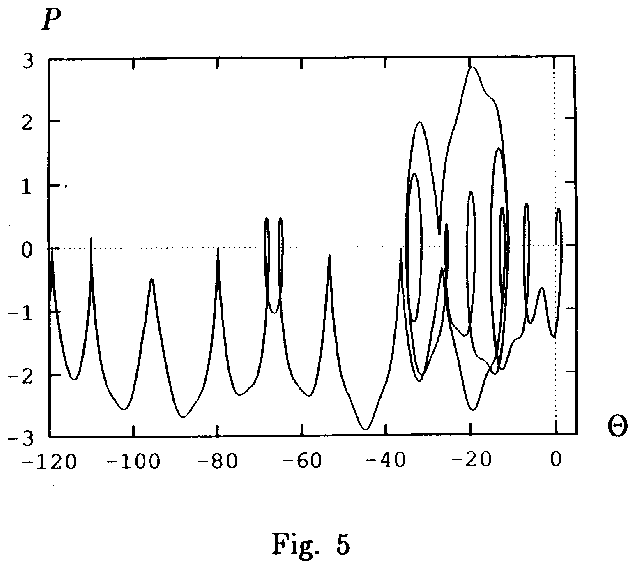}

\epsfxsize=10cm
\hspace{2cm}
\epsfbox{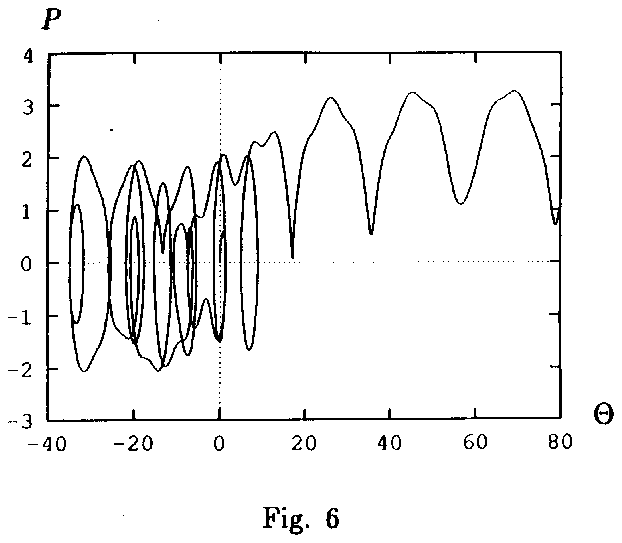}

\epsfxsize=10cm
\hspace{2cm}
\vspace{3cm}
\epsfbox{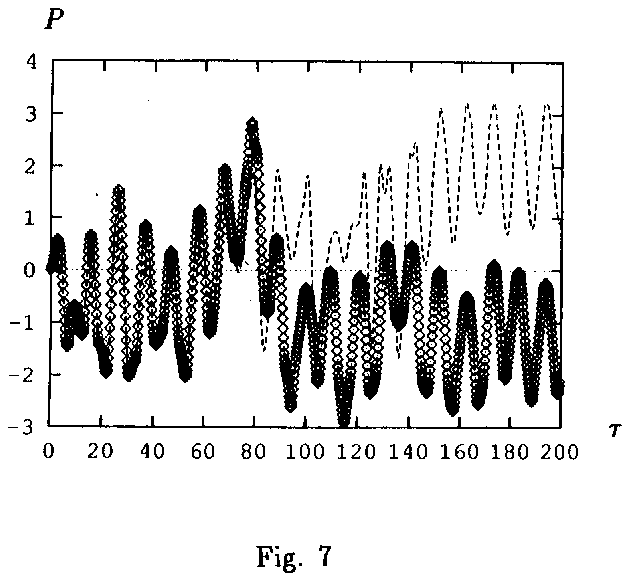}

\epsfxsize=10cm
\hspace{2cm}
\epsfbox{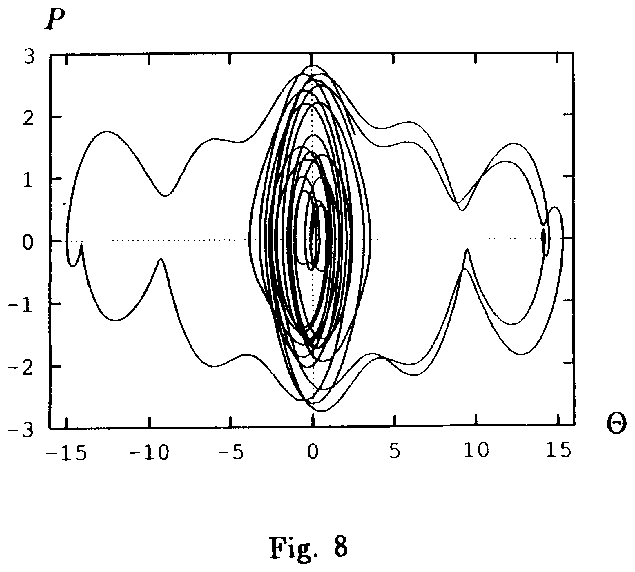}

\epsfxsize=10cm
\hspace{2cm}
\vspace{3cm}
\epsfbox{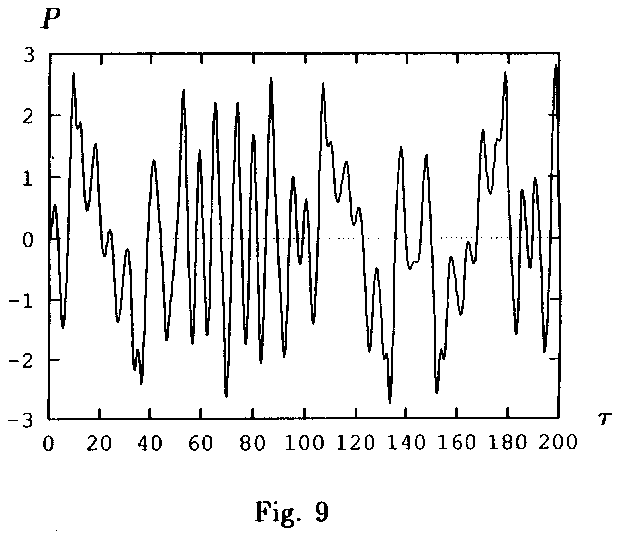}

\epsfxsize=10cm
\hspace{2cm}
\epsfbox{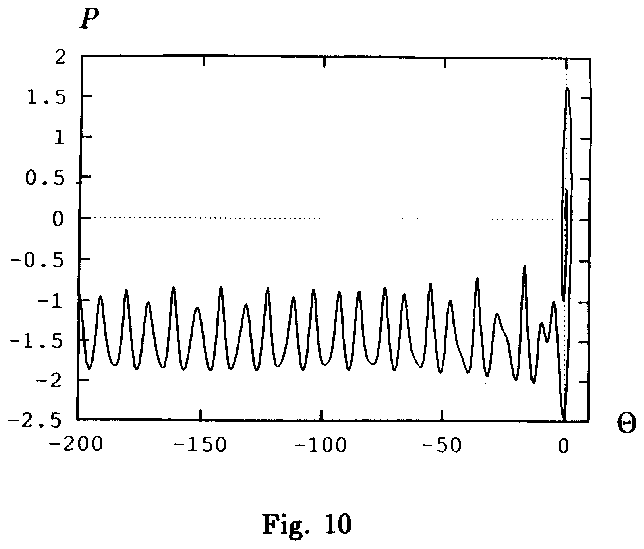}

\epsfxsize=10cm
\hspace{2cm}
\vspace{3cm}
\epsfbox{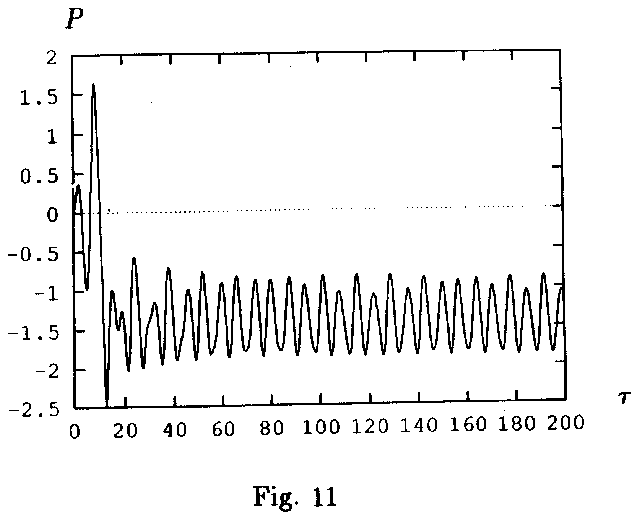}

\epsfxsize=10cm
\hspace{2cm}
\epsfbox{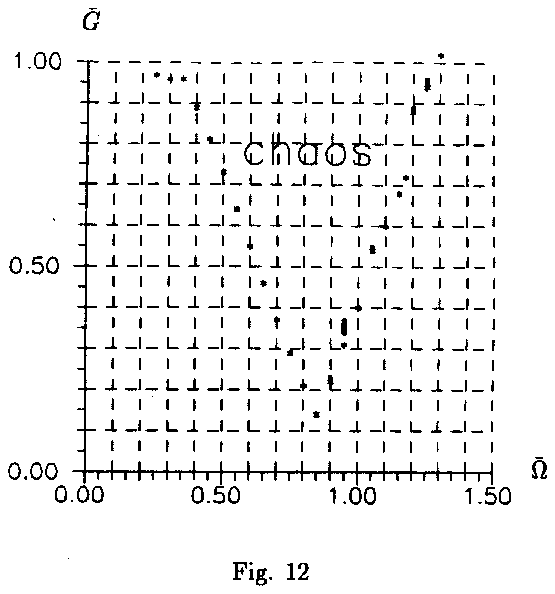}

\epsfxsize=15cm
\hspace{2cm}
\vspace{4cm}
\epsfbox{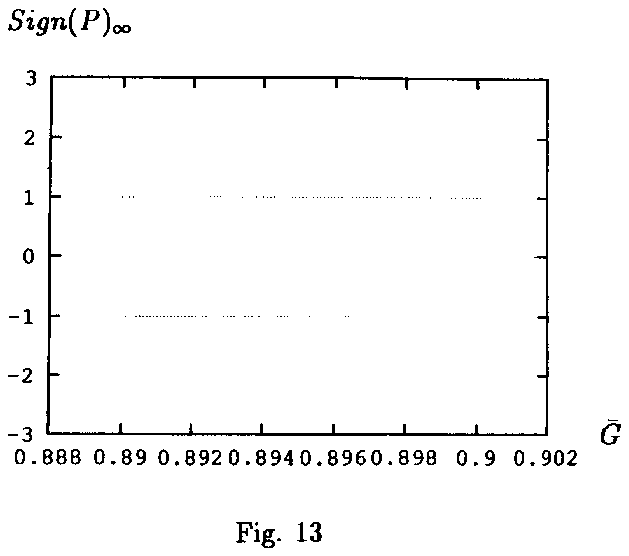}

\end{document}